\documentclass{ws-rv975x65}
\usepackage{ws-rv-van}             
\usepackage{ws-index}             
\makeindex
\newindex{aindx}{adx}{and}{Author Index}       
\renewindex{default}{idx}{ind}{Subject Index}  

\usepackage{makeidx}

\begin{document}

\chapter{From Anderson Localization to Mesoscopic Physics\label{ch1}}

\author[]{Markus B\"{u}ttiker}
\address{University of Geneva, Department of Theoretical Physics,\\
24 Quai E. Ansermet, 1211 Geneva, Switzerland \\
Markus.Buttiker@unige.ch}
\author[M. B\"{u}ttiker and M. Moskalets]{Michael Moskalets}
\address{Department of Metal and Semic. Physics, NTU "Kharkiv Polytechnic Institute", \\
21 Frunze street, 61002 Kharkiv, Ukraine \\
moskalets@kpi.kharkov.ua}
\index[aindx]{B\"{u}ttiker, M.}
\index[aindx]{Moskalets, M.}

\begin{abstract}
In the late seventies an increasing interest in the scaling theory of Anderson localization led to new efforts to understand the conductance of systems which scatter electrons elastically. The conductance and its relation to the scattering matrix emerged as an important subject. This, coupled with the desire to find explicit manifestations of single electron interference, led to the emergence of mesoscopic physics. We review electron transport phenomena which can be expressed elegantly in terms of the scattering matrix. Of particular interest are phenomena which depend not only on transmission probabilities but on both amplitude and phase of scattering matrix elements.
\end{abstract}

\body

\section{Introduction}\label{sec1.1}

Theories and experiments on Anderson localization\cite{Anderson58} have been a vibrant topic of solid state physics for more than five decades. For a long time, it seems, it was the only problem in which disorder and phase coherence were brought together to generate macroscopic effects in electron transport through solids. In particular the scaling theory initially advanced by Thouless\cite{Thouless74} and further developed toward the end of the seventies and early eighties brought the conductance \index{conductance} of disordered systems into focus\cite{Wegner,AALR}. The key argument of the single parameter scaling theory of localization is that the conductance (not conductivity or anything else) is in fact the only parameter that counts. Long after the early work of Landauer\cite{Landauer57,Landauer70}, this emphasis on conductance eventually brought to the forefront the question\cite{LA,Anderson81,Engquist,Azbel,Economou,Fisher,LandauerPLA,BILP}
of a "correct formulation of conductance".  It is on this background that the notion of quantum channels \index{quantum channels} and the scattering matrix \index{scattering!matrix} as a central object of electrical transport theory emerged. A multi-terminal formulation\cite{MBFour,MB1988} of conductance brought much clarity and permitted to connect the scattering theory of electrical conductance to the Onsager theory of irreversible processes. This triggered questions about dynamic fluctuations and led to a very successful theory of current noise in mesoscopic structures\cite{BB}.

\begin{figure}[t]
\centerline{\psfig{file=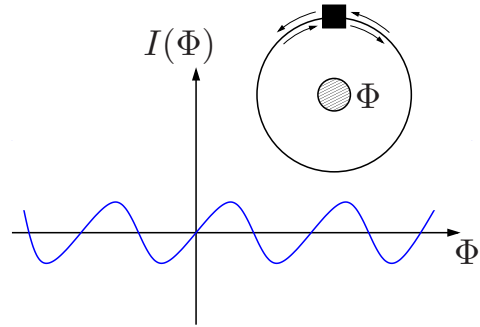,width=6.2cm}}
\caption{Persistent current $I(\Phi)$ as a function of an Aharonov-Bohm flux $\Phi$ of a closed ring with an elastic scatterer (square).}
\label{fig1.1a}
\end{figure}

\begin{figure}[t]
\centerline{\psfig{file=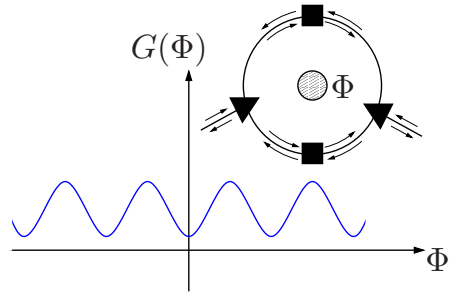,width=6.2cm}}
\caption{Single charge Aharonov-Bohm conductance oscillations of a normal ring with leads. Triangles indicate electron wave beam splitters and elastic scatterers are symbolized by squares.}
\label{fig1.1b}
\end{figure}

The quest to understand the quantum coherent conductance through a disordered region led to more than simply a relation between conductance and transmission. While in the discussions of the localization the calculation of conductance always implied an average over an ensemble of disorder configurations, eventually, the basic question was raised whether we can observed interference effects directly rather than indirectly by considering an ensemble averaged quantity. This brought the specific sample into focus. Early work proposed samples with the shape of a ring with the hole of the ring penetrated by an Aharonov-Bohm flux (see Figs.~ \ref{fig1.1a}, \ref{fig1.1b}). Of interest was the sample specific response to the Aharonov-Bohm flux. The oscillation of the quantity of interest with the period of the single particle flux quantum $h/e$ would be the direct signature of single particle interference.\index{single particle!interference} These expectations led to the prediction \cite{BIL} of persistent currents \index{persistent currents} and Josephson (Bloch) oscillations in closed disordered rings and Aharonov-Bohm oscillations with a period of a single charge flux quantum in rings connected
to leads\cite{GIA,BIA}. We mention here only two recent experiments: one on persistent currents \cite{Harris} and one on Aharonov-Bohm conductance oscillations \cite{Komiyama}. For additional references and a broader account of this development, we refer to Imry's book \cite{Imry}.

The effect of an Aharonov-Bohm flux on the properties of a disordered systems was not entirely novel in mesoscopic physics:\index{mesoscopic!physics} it occurs already in discussions of the Anderson transition. The sensitivity to boundary conditions played a very important part in localization theories. Different transverse boundary conditions for wires are already investigated by Dorokhov\cite{Dorokhov} in the form of an AB-flux along the axes of a cylindrical wire. Nevertheless these discussions were a purely technical means to arrive at an ensemble average conductance. However there is an Aharonov-Bohm effect which shows up even in the ensemble averaged conductance. In the absence of a magnetic field, interference effects of self-intersecting trajectories are not averaged out since there exists a time reversed trajectory which traverses the loop in the opposite direction with exactly the same phase. This leads to enhanced backscattering and, if a magnetic flux is applied to the sample, to oscillations with decaying amplitude and a period in magnetic field determined by the size of the most likely self-intersecting loop. The enhanced backscattering is known as weak localization. The magnetic field oscillations were first discussed by
Altshuler, Aronov and Spivak\cite{AAS}, and were soon  demonstrated in an ingenious experiment by Sharvin and Sharvin\cite{SS} who created a cylinder by coating a glass fiber with a metallic film. This lead to the first observations of an Aharonov-Bohm effect in a dissipative, metallic diffusive conductor.

The concern with the conductance of an elastic scatterer led to a novel approach in which transport quantities are expressed in terms of the scattering matrix. From initial efforts to discuss the conductance in the presence of stationary applied voltages there exist now formulations of transport problems for a wide range of situations.
We will discuss only a few topics, the response to oscillating voltages, which implies a concern with capacitance, quantum pumping, and non-linear dynamic transport leading to a description of single particle emission \index{single particle!emission} from localized states. There are many additional extensions which we will not further discuss here, extensions to superconducting-normal structures, conductors with spin orbit interactions, systems like graphene or topological insulators and superconductors. The message is simply to point to the wider application of scattering theory.
Many of these novel developments concern the variation of the scattering matrix due to a variation of one or several parameters. The scattering matrix can be varied through a displacement of an impurity, a change in a gate voltage or due to a magnetic field. Changes of the scattering matrix lead to transfer of charge out of the scatterer into the contacts or from contacts into the sample.

\section{Charge transfer from the scattering matrix}\label{sec1.2}

Let $S$ be the scattering matrix of a conductor. Its elements are $S_{\alpha\beta}$ where $\alpha = 1,2,..N$ numbers the leads connected to the scattering region. Let there be a small variation in a parameter of the S-matrix that leads to a change in the scattering matrix which we can denote by $d S$.
A variation of the scattering matrix typically implies a transfer of charge from the sample into the leads. An elegant expression for the charge transferred from the scattering region into a contact\cite{avron00,AEGS} $\alpha$ is

\begin{equation}
d{\overline Q}_{\alpha}  = - \frac{e}{2\pi i}\, (dS S^{\dagger})_{\alpha\alpha}\,.
\label{eq1}
\end{equation}
\ \\
This is the emittance \index{emittance} of the sample into contact ${\alpha}$. Similarly, the charge injected into the sample from a lead $\alpha$ is

\begin{equation}
d{\underline Q}_{\alpha}  = - \frac{e}{2\pi i}\, ( S^{\dagger} dS)_{\alpha\alpha}\,.
\label{eq2}
\end{equation}
\ \\
In the absence of a magnetic field injected and emitted charge are the same. In the presence of a magnetic field $B$ the emittance is equal to the injectance \index{injectance} in the reversed magnetic field, $d{\overline Q}_{\alpha} (B)= d{\underline Q}_{\alpha}(-B)$.
These expressions are close relatives of the Wigner-Smith matrix\cite{WS}, a time-shift matrix\cite{AEGS}, scattering matrix expressions of the ac-response of a conductor to a slow potential variation\cite{BTP} and they are now widely used to describe \index{quantum pumping} quantum pumping\cite{avron00,AEGS,Brouwer,PB01,MM01,Moskalets1,ZCM03,AS07,JCSF08}.
Eqs.~(\ref{eq1}, \ref{eq2}) are for non-interacting electrons. To account for the Coulomb interaction $dS$ must be determined self-consistently. Below we discuss how this leads to expressions for the capacitance of mesoscopic conductors\cite{MBcap}. \index{mesoscopic!conductors}

For a single channel, two terminal conductor we can parameterize the scattering matrix as follows:

\begin{equation}
S= e^{i\gamma}\, \left(\begin{array}{cc} e^{i\alpha}\, \cos \theta  &\, i\,e^{-i\phi}\, \sin \theta \\
i\,e^{i\phi}\ \sin \theta &\, e^{-i\alpha}\, \cos \theta
\end{array}\right) \,.
\label{eq3}
\end{equation}
\ \\
A variation of $\gamma, \alpha$ and $\phi$ gives each rise to a separate response and allows for an elementary interpretation\cite{AEGS}. The resulting emittance into the left contact is:

\begin{equation}
d{\overline Q}_{L}  = \frac{e}{2\pi} (- \cos^2 \theta\,d\alpha + \sin^2 \theta\, d\phi -d\gamma )
\label{eq4}
\end{equation}
\ \\
and into the right contact it is

\begin{equation}
d{\overline Q}_{R}  = \frac{e}{2\pi} (\cos^2 \theta\,d\alpha - \sin^2 \theta\, d\phi - d\gamma )
\label{eq5}
\end{equation}
\ \\
\indent
The total charge emitted by the sample is $dQ = d{\overline Q}_R + d{\overline Q}_L = -e\, d\gamma /\pi $. This is nothing but the Friedel sum rule\cite{Friedel} which relates the variation of the phase $\gamma$ to the charge of the scatterer (Krein formula in mathematics).

The phase $\alpha$ describes the asymmetry in reflection at the scatterer of carriers incident from the left as compared to those incident from the right. A displacement $dx$ of the scatterer to the right changes $\alpha$ by $d\alpha = 2k_F dx$. It absorbs charge from the left lead and emits it into the right lead. Note that the transmission can be completely suppressed\cite{AEGS} and the scatterer than acts as a "snow plow".

The phase $\phi$ can be non-vanishing only if time-reversal symmetry is broken, that is, if a magnetic field acts on the conductor. Therefore we can suppose that $\phi$ is the result of a vector potential,
$\phi = - 2\pi \int dx A/\Phi_0$ where $\Phi_0 $ is the single charge flux quantum. Suppose that $A$ is generated by an Aharonov-Bohm flux $\Phi$ which depends linearly on time.  The electric field generated by the flux induces a voltage $V$ between the reservoirs connected to the sample. The phase $\phi$ increases as $\phi = -V t/\Phi_0$ with $V$ the voltage induced across the sample. We then have\cite{AEGS,Cohen} $d{\overline Q}_{\alpha}/dt = - e \sin^2 \theta\, d\phi/dt = (e^2 /h) T V$ which with the transmission probability $T = \sin^2 \theta $ is the Landauer conductance $G = (e^2 /h) T$.

The phase $\gamma$ multiples all channels. It is a phase that depends on what we consider the scatterer and the outside regions. In one dimension, on the line, phases of the scattering matrix are defined from two points to the right and left of the scatterer. (More generally scattering phases depend on the choice of cross-sections).
If we extend the region that we attribute to the scatterer by $dx$ on both sides of the scatterer the phase $\gamma$ changes by $d\gamma = 2k_F dx$. This operation removes charges from the leads and we have
$d{\overline Q}_L = d{\overline Q}_R = -e\, k_F dx /\pi $.

A variation of the parameter $\theta$ which corresponds to a variation of the transmission probability has no effect on the emittance if all phases are kept fixed.

\section{The Wigner-Smith delay time and energy shift matrix}\label{sec1.3}

There are two close relatives of the expressions for charge transfer given by Eqs.~(\ref{eq1}) and (\ref{eq2}).
With both the emittance Eqs.~(\ref{eq1}) and the injectance Eqs.~(\ref{eq2}) we can associate a Wigner-Smith delay time matrix \cite{WS}. In the first case we focus on the time-delay in the outgoing channel regardless of the input channel and in the second case we specify the incoming channel but sum over all outgoing channels. Multiplying the emittance per unit energy with $\hbar/e$ gives the time delay matrix \index{matrix!time delay}

\begin{equation}
{\overline {\cal T}} = \frac{\hbar}{i}\,\frac{\partial { S}}{\partial E} { S}^{\dagger}\,.
\label{eq6}
\end{equation}
\ \\
The Wigner phase delay times are the diagonal elements of this matrix.
The off-diagonal elements also have physical significance and going back to charge we can related them to charge density fluctuations\cite{MBmath,Pedersen}.  There is a dual to the Wigner-Smith matrix which Avron et al.\cite{AEGS} call the energy shift matrix.\index{matrix!energy shift} It is important once the scattering matrix is taken to be time-dependent. The energy shift matrix is

\begin{equation}
{\overline {\cal E}} = i \hbar\, \frac{\partial{ S}}{\partial t} { S}^{\dagger}\,.
\label{eq7}
\end{equation}
\ \\
The two matrices do not commute, reflecting the time-energy uncertainty.
Since they are matrices, we can define two commutators (Poisson brackets) depending on the order in which they are taken.
These commutators play an important role whenever there is a time-variation of the scattering matrix.
One of them,

\begin{equation}
{\cal P}\left\{S, S^{\dag} \right\} = \frac{i }{\hbar } \left(
{\overline {\cal T}}\, {\overline {\cal E}} - {\overline {\cal E}}\, {\overline {\cal T}} \right) =  i \hbar\, \left( \frac{\partial{ S}}{\partial t} \frac{\partial{ S}^{\dagger}}{\partial E} - \frac{\partial{ S}}{dE}  \frac{\partial{ S}^{\dagger}}{\partial t} \right) .
\label{eq8}
\end{equation}
\ \\
defines the spectral densities of currents \index{spectral densities of currents} generated by a parametric variation of the scattering matrix \cite{AEGS, Moskalets2}.
The other order, ${\cal P}\left\{S^{\dag}, S \right\}$, defines corrections to the scattering matrix proportional to the frequency $\omega$ with which parameters of the scattering matrix are varied.\cite{Moskalets2,MBmag05}

\section{The internal response}\label{sec1.4}

The energy shift matrix multiplied by $e/h$ defines a matrix of currents. The diagonal elements $\alpha \alpha$ are the currents generated in contact $\alpha$ in response to a temporal variation of the S-matrix,

\begin{equation}
dI_{\alpha} = \frac{e}{2\pi i}\, \left(\frac{\partial { S}}{\partial t} { S}^{\dagger}\right)_{\alpha\alpha}\,.
\label{eq9}
\end{equation}
\ \\
Assuming that the scattering matrix ${ S}$ is a function of the electrostatic potential $U(t)$
we can write

\begin{equation}
{dI}_{\alpha}= \frac{e}{2\pi i}\, \left(\frac{\partial { S}}{\partial U} { S}^{\dagger}\right)_{\alpha\alpha}\,
\frac{dU}{dt}
\label{eq10}
\end{equation}
\ \\
or if we take the Fourier transform

\begin{equation}
dI_{\alpha}(\omega)= - ie\omega\, \frac{1}{2\pi i}\, \left(\frac{\partial { S}}{\partial U} { S}^{\dagger}\right)_{\alpha\alpha}\,\,
dU({\omega}) \,.
\label{eq11}
\end{equation}
\ \\
Eq.~(\ref{eq11}) gives the current in contact $\alpha$ in response to a potential oscillating over the entire region of the scatterer.
The potential $U$ typically is a function not only of time but also of space $r$. This can be taken into account by testing the scattering matrix with respect to a small localized potential at every point $r$ in the sample and integrating the result over the entire scattering region,

\begin{equation}
dI_{\alpha}(\omega)= - ie\omega\, \frac{1}{2\pi i}\, \int dr^{3} \left(\frac{\delta { S}}{\delta U(r)}  { S}^{\dagger}\right)_{\alpha\alpha}\,\,\delta U(r, \omega)\,.
\label{eq12}
\end{equation}
\ \\
This is the result obtained by one of the authors in collaboration with H.~Thomas and A.~Pr\^{e}tre \cite{BTP}. Eq.~(\ref{eq12}) is a zero temperature result. It depends on the scattering matrix only at the Fermi energy.

\subsection{Quantum pumping}\label{sec1.5}

For a sinusoidal potential the time average current vanishes. However, if $U(r,t)$ cannot be written as a product of a time-dependent function times a spatial function, but effectively varies the potential of the conductor at several points and out of phase, the current integrated over a period ${\cal T}$ does not average to zero, and Eq.~(\ref{eq9}) describes a pump current\cite{Brouwer}

\begin{equation}
{I}_{dc,\alpha} = \frac{e}{2\pi i {\cal T} }\, \oint_{\cal L} \left( d{ S}\, { S}^{\dagger}\right)_{\alpha\alpha}\,,
\label{eq13}
\end{equation}
\ \\
where ${\cal L}$ is the "pump path".
It is often stated that pumping needs at least two parameters which oscillate out of phase. These two parameters describe the pumping path ${\cal L}$ in the parameter space.
A two parameter formulation of pumping based on the concept of emittance has been given by Brouwer\cite{Brouwer}.
In terms of two parameters
$X_\mathrm{1}(t) = X_\mathrm{1}\cos(\omega t + \varphi_\mathrm{1})$ and
$X_\mathrm{2}(t) = X_\mathrm{2}\cos(\omega t + \varphi_\mathrm{2} )$ Brouwer obtained for weak pumping

\begin{equation}
\label{eq14}
I_\mathrm{dc,\alpha} = \frac{e\omega\sin(\varDelta\varphi)
X_\mathrm{1}X_\mathrm{2}}{2\pi}
\sum\limits_{\beta=1}^{N_\mathrm{r}}
\Im\left(
\frac{\partial S^{*}_{\alpha\beta}}{\partial X_\mathrm{1}}
\frac{\partial S_{\alpha\beta}}{\partial X_\mathrm{2}}
\right)_\mathrm{X_\mathrm{1}=0,X_\mathrm{2}=0}\,,
\end{equation}
\ \\
where $\varDelta\varphi\equiv\varphi_\mathrm{1}
-\varphi_\mathrm{2}\neq 0$ is the phase lag. However, single parameter pumps are also possible: A typical example is an Archimedes screw where we turn only one handle to generate a pumped current\cite{QZ09, OZL09}.
If the rotation is with a constant speed, $\chi = \omega t$, and the scattering matrix is periodic in a rotation angle, $S(\chi) = S(\chi + 2\pi)$, then accordingly to Eq.~(\ref{eq13}) the dc current is not zero, $I_{dc,\alpha} \ne 0$, if the Fourier expansion for a diagonal element $(\partial S/\partial\chi S^{\dag})_{\alpha\alpha}$ includes a constant term.

Equation (\ref{eq13}) is valid for both weak and strong pumping but requires that the pump is driven slowly.
At higher pumping frequency or pulsed excitation the generated currents can not be described merely via a temporal variation of a  scattering matrix dependent on a single energy.
In contrast the scattering matrix dependent on two energies\cite{MBstrong02,AM06} or, equivalently, two times\cite{PB03,Vavilov05} has to be used.

\subsection{Pumping in insulators and metals}\label{sec1.6}

Thouless investigated pumping in band insulators\cite{Thouless}. A specific example considered is a potential that is periodic in space with lattice constant $a$ and moves with a constant velocity $v$ and is (in one dimension) of the form $U(x,t) = U(x-v t)$. Depending on the number $n$ of bands filled the charge moved by advancing the potential by a period $a$ in space or by ${\cal T} = a/v$ in time is quantized and given by a Chern number $n e$. This formulation assumes that the Fermi energy lies in a spectral gap.
On the other hand in gap-less Anderson insulators quantized pumping \index{quantized!pumping} is also possible.
It is due to tunneling resonances through the sample\cite{COMN07}.

Thouless's result seems hard to reconcile with a pump formula that is based explicitly only on the properties of the scattering matrix at the Fermi energy.
However, Graf and Ortelli\cite{Graf} have shown that by taking a periodic
lattice of some finite length L and coupling it on either side to metallic contacts, the scattering approach converges, in the limit of a very long lattice,
to the quantized pump current of Thouless.
The point is of course, that as soon as the lattice is taken to be finite, there are exponentially decaying states which penetrate from the contacts into the sample. Considering the scattering matrix due to these states gives in the long length limit the same result as found from Thouless's approach. While the one-channel case permits to solve this problem by direct calculation, the many channel case requires more thought\cite{Braeunlich}.

\subsection{Scattering formulation of ac-response}\label{sec1.7}

Consider a conductor connected to multiple contacts\cite{PTB,BTP} and suppose that the voltages $V_{\beta}(\omega)$ at the contacts oscillate with frequency $\omega$. In addition the conductor is capacitively coupled to a gate at voltage $V_g$. The coupling is described in terms of a geometrical capacitance $C$.\index{capacitance!geometrical} The currents $dI_{\alpha}(\omega)$ are related to the voltages by a dynamical conductance matrix \index{conductance!dynamical} $G_{\alpha\beta}(\omega)$. In linear response the total current has in general two contributions: a current that is solely the response to the external voltage applied to the contact defines a conductance $G^{ext}_{\alpha\beta}(\omega)$ and a current that arises because the application of external voltages charges the sample which leads through interaction to an internal potential \index{internal potential} $U(t)$,

\begin{equation}
\label{eq15}
dI_{\alpha}(\omega) = \sum_{\beta}\, G^{ext}_{\alpha\beta}\,  dV_{\beta}\, + i \omega\,  \Pi_{\alpha}\, dU \,.
\end{equation}
\ \\
Both response functions $G^{ext}_{\alpha\beta}$ and $\Pi_{\alpha}$ remain to be determined. The current at the gate is

\begin{equation}
\label{eq16}
I_{g}(\omega) = -i \omega\, C\,(dV_{g} - dU )  \,.
\end{equation}
\ \\
The internal potential is found self-consistently by the requirement that potentials are defined only up to a constant (an overall shift in potentials does not change a physical quantity). This implies

\begin{equation}
\label{eq17}
i \omega\,  \Pi_{\alpha} = -  \sum_{\beta} G^{ext}_{\alpha\beta}\,.
\end{equation}
\ \\
The external response \index{response!external} can be expressed in terms of the scattering matrix and the Fermi function of the contacts\cite{PTB}

\begin{equation}
\label{eq18}
G^{ext}_{\alpha\beta}(\omega) = \frac{e^{2}}{h} \int dE\,    { Tr}[{\bf 1}_{\alpha}\delta_{\alpha\beta} -S^{\dagger}_{\alpha\beta} (E) S_{\alpha\beta}(E +\hbar \omega )] \frac{f_{\beta} (E) -  f_{\beta}(E+\hbar \omega )}{\hbar\omega}\,.
\end{equation}
\ \\
Here we now consider leads with several transverse channels. The trace $Tr$ is a sum over transverse channels in the leads ${\alpha}$ and ${\beta}$.
Using the above and requiring that current is conserved determines the internal potential. Eliminating the internal potential leads to a total conductance\cite{PTB}

\begin{equation}
\label{eq19}
G_{\alpha\beta} =  G^{ext}_{\alpha\beta} + \frac{ \sum_{\gamma} G^{ext}_{\alpha\gamma} \sum_{\delta} G^{ext}_{\delta\beta}} {i \omega C - \sum_{\gamma\delta} G^{ext}_{\gamma\delta}}\,.
\end{equation}
\ \\
The second term is the internal response \index{response!internal} due to the self-consistent potential. It depends on the capacitance $C$.

A voltage at contact $\beta$ of the conductor does, in the absence of interactions, not induce a response
at the gate. Thus if $\alpha = g$ in Eq.~(\ref{eq19}) the external response $G^{ext}_{g \beta}$ vanishes and $\sum_{\delta} G^{ext}_{\delta\beta}$ has to be replaced by $-i\omega C$. Similarly, an oscillating gate voltage does not generate an external response  $G^{ext}_{\alpha g}$ but due to capacitive coupling generates a current at contact $\alpha$ due to second term of Eq.~(\ref{eq19}) in which $\sum_{\delta} G^{ext}_{\delta g}$ is replaced by $-i\omega C$.
The rows and columns of the conductance matrix add up to zero.
Next we now discuss the low frequency limit of these results and show the role played by the emittances and injectances.

\subsection{Capacitance, emittances, partial density of states}\label{sec1.8}

An expansion of the external conductance to first order in $\omega$ leads to

\begin{equation}
G^{ext}_{\alpha\beta}(\omega) = G_{\alpha \beta} -i \omega e^{2} dN_{\alpha \beta}/dE +...\,\,
\label{eq20}
\end{equation}
\ \\
where the first term is the dc-conductance and

\begin{equation}
\frac{dN_{\alpha\beta}}{dE} = \frac{1}{4\pi i}\,{ Tr} \left\{ \frac{dS_{\alpha\beta}}{dE}\,S^{\dagger}_{\alpha\beta} - S_{\alpha\beta}\,\frac{dS^{\dagger}_{\alpha\beta}}{dE} \right\}\,\,
\label{eq21}
\end{equation}
\ \\
is a partial density of states.\index{density of states!partial} In contrast to the emittance for which the contact into which carriers are emitted are specified or the injectance for which the contact from which carriers are injected is specified, here both the contact $\beta$ from which carriers are injected and the contact $\alpha$ into which they are emitted are specified. For the injectance we have a pre-selection, for the emittance a post-selection, but for the partial density of states we have both a pre- and post-selection of the contacts. The sum of partial density of states over the second index is the emittance $d{\overline N}_{\alpha}/dE = \sum_{\beta} dN_{\alpha\beta}/dE$. The sum over the first index of the partial density of states is the
injectance $d{\underline N}_{\alpha}/dE = \sum_{\beta} dN_{\beta\alpha}/dE$.
The sum over both indices is the total density of states $ dN/dE =
\sum_{\alpha\beta}\, dN_{\alpha\beta}\, dE $
Using this we find for the screened emittance\cite{PTB} \index{emittance!screened}

\begin{equation}
E_{\alpha\beta} = e^{2}\left[ \frac{dN_{\alpha \beta}}{dE} - \frac{d{\overline N}_{\alpha}}{dE} \frac{e^{2}}{C+ e^{2}dN/dE} \frac{d{\underline N}_{\beta}}{dE} \right]
\,\,
\label{eq22}
\end{equation}
\ \\
A positive diagonal term $E_{\alpha\alpha} > 0$ (a negative off-diagonal term, $E_{\alpha\beta}  < 0$) signals a capacitive response. \index{response!capacitive} A negative diagonal term $E_{\alpha\alpha} < 0$ (a positive off-diagonal term, $E_{\alpha\beta} > 0$) describes an inductive response. \index{response!inductive} We have not taken into account magnetic fields and hence the inductive response is of purely kinetic origin. The main point for our discussion is the physical relevance of the different density of states which is brought out by Eq.~(\ref{eq22}). The Coulomb term illustrates that an additional injected charge generates an electrical potential which in turn emits charge.

\subsection{AC response of a localized state}\label{sec1.9}

\begin{figure}[t]
\centerline{\psfig{file=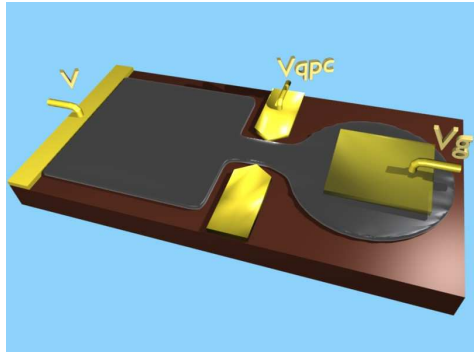,width=6.2cm}}
\caption{Mesoscopic capacitor. A cavity separated via a quantum point contact at voltage $V_{QPC}$ from a lead which is in turn coupled to a metallic contact at a voltage $V$. The cavity is capacitively coupled to a gate $V_g$.}
\label{fig1.1}
\end{figure}

\begin{figure}[t]
\centerline{\psfig{file=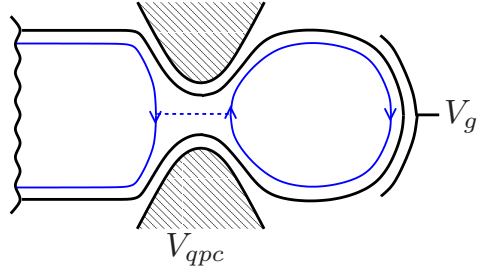,width=6.2cm}}
\caption{A localized state (edge state) coupled via a quantum point contact to a lead and reservoir}
\label{fig1.2}
\end{figure}

Modern sample fabrication permits to investigate a single localized state coupled only via one contact of arbitrary transmission to a metallic contact (see Fig.~\ref{fig1.1} and Fig.~\ref{fig1.2}). Such a structure can be viewed as a mesoscopic version of a capacitor. \index{mesoscopic!capacitor} A particular realization consists of a quantum point contact (QPC) which determines the coupling to the inside of a cavity. In the presence of a high magnetic field there is a single edge state coupled to the contact (see Fig.~\ref{fig1.2}). The dc-conductance of such a localized state is zero since every carrier that enters the cavity is after some time reflected. However by coupling the cavity to a (macroscopic) gate an ac voltage can be applied and the ac-conductance of the localized state can be investigated.
For a macroscopic capacitor the contact of the cavity to the metallic reservoir would act as a resistor and the low frequency conductance would be
\begin{equation}
G(\omega) = -i \omega C + \omega^{2} C^{2} R  +..   \,\,,
\label{eq23}
\end{equation}
\ \\
with $C$ the geometrical capacitance and $R$ the series resistance. A mesoscopic capacitor has a response of the same form but now with quantum corrections to the capacitance and the resistance,

\begin{equation}
G(\omega) = -i \omega C_{\mu} + \omega^{2} C^{2}_{\mu} R_q  +..   \,\,.
\label{eq24}
\end{equation}
\ \\
Here $C_{\mu}$ is an electrochemical capacitance \index{capacitance!electrochemical} and $R_q$ is the charge relaxation resistance.\index{resistance!charge relaxation}
Specializing Eq.~(\ref{eq15}) to a single contact, yields\cite{BTPPLA}

\begin{equation}
C^{-1}_{\mu} = C^{-1} +\left[ e^{2} \nu(\mu) \right]^{-1} \,,
\label{eq25}
\end{equation}
\ \\
where $\nu(E) \equiv dN/dE = (1/2\pi)\, Tr[S^{\dagger}dS/dE]$ is the density of states of the cavity. \index{density of states}
For weak interaction $C$ is large and $C^{-1}_{\mu}$
is entirely determined by the density of states. For strong interaction $C$ is small and the density of states provides a small correction to the geometrical capacitance.

The charge relaxation resistance is a particularly interesting transport coefficient. From Eq.~(\ref{eq15}) we find\cite{BTPPLA,PTB}

\begin{equation}
R_q  = \frac{h}{2e^{2}} \frac{Tr[dS^{\dagger}/dE\,dS/dE] }{(|Tr[S^{\dagger}dS/dE]|)^{2}}\,\,.
\end{equation}
\ \\
For the single contact considered here the injectance and emittance are identical.
In the basis in which the scattering matrix is diagonal its elements are $e^{i\phi_n} $ $n =1, 2,..$ and we can express the matrix $d{\overline N}/dE$ in terms of energy
derivatives of phases, the Wigner phase delays,

\begin{equation}
\left[\frac{dN }{dE}\right] = \frac{1}{2\pi i}\,
Tr\left[\frac{\partial S}{\partial E}\, S^{\dagger} \right] = \frac{1}{2\pi}\, \sum_{n}\, \frac{\partial\phi_{n}}{\partial E}\,\,,
\nonumber
\end{equation}
\begin{equation}
\frac{1}{(2\pi)^{2} }
Tr\left[\frac{\partial S^{\dagger}}{\partial E} \frac{\partial S}{\partial E}\right] = \frac{1}{(2\pi)^{2} } \sum_{n} \left(\frac{\partial\phi_{n}}{\partial E}\right)^{2}\,\,,
\nonumber
\end{equation}
\ \\
and thus

\begin{equation}
R_q  = \frac{h}{2e^{2}} \frac{\sum_{n}( \partial\phi_{n}/\partial E)^{2} }{(\sum_{n}\partial\phi_{n}/\partial E)^{2}}\,\,.
\label{eq28}
\end{equation}
\ \\
Interestingly this is now a resistance that is not expressed in terms of transmission probabilities but in terms of derivatives of phases. It is the ratio of the mean square time of carriers in the cavity divided
by the mean delay time squared. For a single quantum channel this two quantities are equal and

\begin{equation}
R_q  = \frac{h}{2e^{2}} \,\,
\label{eq29}
\end{equation}
\ \\
is universal\cite{BTPPLA,PTB}, i.e. independent of the scattering properties of the quantum channel!! This derivation assumes that the quantum coherence can be maintained long compared to the time it takes a carrier to be reflected from the cavity.
If the phase breaking time becomes comparable to the dwell time,
theory predicts deviations from the above quantized resistance \index{quantized!resistance} value \cite{Nigg1}.
Note the factor $2$ in Eq.~(\ref{eq29}) is not a consequence of spin: the quantization at half of a resistance value is for a single spin channel.

An experiment with a setup as shown in Fig.~\ref{fig1.1} has been carried out by Gabelli et al.\cite{gabelli06} and has provided good evidence of the resistance quantization of $R_q$.
In the experiment \cite{gabelli06} the quantum point contact is modeled by a scattering matrix with transmission amplitude $t$ and reflection amplitude $r$ (taken to be real) and a phase $\phi$ which is accumulated by an electron moving along the edge state in the cavity. The amplitude of the current incident from the metallic contact on the QPC $a$, the amplitude of the transmitted wave leaving the QPC $b$, which after completing a revolution is  $\exp(i \phi ) b$, and the amplitude of the current leaving the QPC $s a$, with $s$ the scattering matrix element, are related by

\begin{equation}
\left(\begin{array}{c} sa\\b\end{array}\right) \,\,\,
= \left(\begin{array}{cc} r & -t\\
t & r \end{array}\right) \,
\left(\begin{array}{c} a\\
\exp(i \phi ) b \end{array}\right) \,.
\label{eq30}
\end{equation}
\ \\
This determines the scattering matrix element

\begin{equation}
s (\epsilon) = - e^{-i\phi} \frac{r-e^{i\phi}}{r-e^{-i\phi}}
\label{eq31}
\end{equation}
\ \\
and gives a density of states

\begin{equation}
\nu(E) = \frac{1}{2\pi i}
s^{\dagger} \frac{\partial s}{\partial E} = \frac{1}{2\pi i}\,
s^{\dagger}\, \frac{ds}{d\phi}\,\frac{\partial\phi}{\partial E}\,=
\frac{1}{2\pi} \frac{\partial\phi}{\partial E}
\frac{1-r^{2}}{1-2r cos(\phi)+ r^{2}}\,
\label{eq32}
\end{equation}
\ \\
Since the scattering matrix is needed only around the Fermi energy it is reasonable to assume that the phase is a linear function of energy

\begin{equation}
\phi  = 2 \pi E /\Delta \,,
\label{eq33}
\end{equation}
\ \\
where $\Delta$ is the level spacing. For small transmission the density of states is sharply peaked at the energies for which $\phi$ is a multiple of $2 \pi$. In the experiment the capacitance $C_{\mu}$ and $R_q$ are investigated as function of the voltage applied to the QPC. This dependence enters through the specification of the transmission probability of the QPC,

\begin{equation}
\label{eq34}
T= t^{2}  = 1/\left(1+ exp(-(V_{QPC} -V_{0})/\varDelta V_{0})\right)\,\,.
\end{equation}
\ \\
Here $V_{0}$ determines the voltage at which the QPC is half open and $\varDelta V_{0}$ determines the voltage scale over which the QPC opens.

Eq.~(\ref{eq24}) and Eqs.~(\ref{eq29}) - (\ref{eq33}) permit a very detailed description of the experimental data \cite{gabelli06}.
Indeed the experiment is in surprisingly good agreement with scattering theory.\index{scattering!theory} The main reason for this agreement is the weak interaction which in the experimental arrangement is generated by a top gate that screens the cavity.
In the presence of side gates interactions might become more important and it is then of interest to provide theories which go beyond the random phase approximation discussed above. We refer the reader to
recent discussions \cite{Nigg2,BN07,Imry2,Rodionov,LeHur,Martin,Splettstoesser}.

\section{Nonlinear AC response of a localized state}\label{sec1.10}

An experiment by F\`{e}ve et al.\cite{feve07}, with the same setup as shown in Fig.~\ref{fig1.1}, demonstrated {\it quantized particle emission} \index{quantized!particle emission} from the localized state when the capacitor was subject to a large periodic potential modulation, $U(t) = U(t + {\cal T})$, comparable with $\Delta$.
F\`{e}ve et al. used either sinusoidal or pulsed modulations.
We consider the former type of modulation.
The theory for the latter one can be found in Ref.~\refcite{MSB08}.

Let the gate $V_{g}$, see Fig.~\ref{fig1.2}, induce a sinusoidal potential on the capacitor, $U(t) = U_{0} + U_{1}\, \cos\left( \omega t + \varphi \right)$.
If $U_{1} \sim \Delta$ then we can not use a linear response theory and instead of conductance $G(\omega)$ we have to consider directly the current $I(t)$ that is generated.
At low frequency, $\omega \to 0$, we can expand $I(t)$ in powers of $\omega$ in the same way as we do in Eq.~(\ref{eq24}).
However, now the density of states depends on the potential $U(t)$ and thus on time, $\nu(t,E) = \nu[U(t),E]$, where $\nu[U(t),E]$ is given in Eq.~(\ref{eq32}) with $\phi(E)$ being replaced by $\phi[U(t),E] = 2\pi[E  - eU(t)]/\Delta$.
At zero temperature and assuming the geometrical capacitance $C \to \infty$ we obtain up to $\omega^2$ terms,\cite{MSB08}

\begin{equation}
I(t) = C_{\partial}\, \frac{dU }{dt } - R_{\partial} C_{\partial}\, \frac{\partial }{\partial t } \left[ C_{\partial} \frac{dU }{dt } \right]\,,
\label{eq34_1}
\end{equation}
\ \\
\noindent
\ \\
with a differential capacitance \index{capacitance!differential}

\begin{equation}
C_{\partial} = e^2 \nu(t,\mu)\,,
\label{eq35}
\end{equation}
\ \\
\noindent
and a differential resistance \index{resistance!differential}

\begin{equation}
R_{\partial} = \frac{h }{2e^2 }\left\{ 1 + \frac{\frac{\partial \nu(t,\mu) }{\partial t }\, \frac{dU }{dt } }{\frac{\partial }{\partial t } \left[ \nu(t,\mu)\, \frac{dU }{dt } \right]  } \right\}\,.
\label{eq36}
\end{equation}
\ \\
\noindent
We see that in the non-linear regime the dissipative current through the capacitor is defined by the resistance $R_{\partial}$ which, even in a single-channel case, unlike the linear response resistance $R_{q}$, Eq.~(\ref{eq29}), is not universal and depends on the parameters of the localized state, on the strength of coupling to the extended state, and on the driving potential $U(t)$.

Remarkably, in both linear and non-linear regimes the charge relaxation resistance quantum $R_{q} = h/(2e^2)$ \index{resistance!charge relaxation!quantum} defines the Joule heat which is found in the leading order in $\omega$ to dissipate with a rate,\cite{MB09}

\begin{equation}
I_{E} = R_{q} \left \langle  I^2 \right \rangle =  \frac{h }{2e^{2} }\, \int\limits_{0}^{\cal T} \frac{dt}{\cal T}\, I^2(t)\,.
\label{eq37}
\end{equation}
\ \\
\noindent
Note $I_{E}$ is calculated as a work done by the potential $U(t)$ under the current $I(t)$ during the period ${\cal T}$.

\section{Quantized charge emission from a localized state}\label{sec1.11}

If the potential of a capacitor is varied in time, then the energy of the localized state (LS) changes.
With increasing potential energy $e U(t)$ the level of the LS can rises above the Fermi level and an electron occupying this level leaves the capacitor.
When $e U(t)$ decreases, the empty LS can sink below the Fermi level and an electron enters the capacitor leaving a hole in the stream of electrons in the linear edge state.
Therefore, under periodic variation of a potential the localized state can emit non-equilibrium electrons and holes propagating away from the LS within the edge state which acts here similar to a waveguide.

Consider the localized state weakly coupled to the extended state.
In our model this means that the transparency of a QPC is small, $T \to 0$.
Then the generated AC current consists of positive and negative peaks with width $\Gamma_{\tau} \sim T/(2\pi \omega)$ corresponding to emitted electrons and holes, respectively.
Assume the amplitude $U_{1}$ of an oscillating potential to be chosen such that during the period only one quantum level $E_{n}$ crosses the Fermi level.
The time of crossing $t_{0}$ is defined by the condition $\phi(t_{0}) = 0 \mod 2\pi$.
There are two times of crossing.
At time $t_{0}^{(-)}$, when the level rises above the Fermi level, an electron is emitted, and at time $t_{0}^{(+)}$, when the level sinks below the Fermi level, a hole is emitted.
If without the potential $U$ the level $E_{n}$ aligns with the Fermi energy $\mu$, then the times of crossing are defined by $U\left(t_{0}^{(\mp)} \right) = 0$.
For $|eU_{0}| < \Delta/2$ and $|eU_{0}| < |eU_{1}| < \Delta - |eU_{0}|$ we find the emission times, $t_{0}^{(\mp)} = \mp t_{0}^{(0)} - \varphi/\omega$, where $\omega t_{0}^{(0)} = \arccos\left(-  U_{0}/ U_{1} \right)$. Here $\varphi$ is the phase lag of the oscillating potential introduced above.
With these definitions we find the scattering amplitude, Eq.~(\ref{eq31}), for electrons with the Fermi energy

\begin{equation}
s(t,\mu) = e^{i\theta_{r}} \left\{
\begin{array}{ll}
\dfrac{t - t_{0}^{(+)} - i \Gamma_{\tau} }{t - t_{0}^{(+)} + i \Gamma_{\tau} }\,, & \left|t - t_{0}^{(+)} \right| \lesssim \Gamma_{\tau} \,, \\
\ \\
\dfrac{t - t_{0}^{(-)} + i \Gamma_{\tau} }{t - t_{0}^{(-)} - i \Gamma_{\tau} }\,, & \left|t - t_{0}^{(-)} \right| \lesssim \Gamma_{\tau} \,, \\
\ \\
1\,, & \left|t - t_{0}^{(\mp)} \right| \gg \Gamma_{\tau} \,,
\end{array}
\right.
\label{eq38}
\end{equation}
\ \\
\noindent
where $\omega\Gamma_{\tau} = T\Delta/\left( 4 \pi |e| \sqrt{U_{1}^2 - U_{0}^2}  \right)$ and $0 < t < {\cal T}$.
The density of states,

\begin{equation}
\nu(t,\mu) = \dfrac{4}{\Delta T} \left\{ \dfrac{\Gamma_{\tau}^{2} }{ \left(t - t_{0}^{(-)} \right)^{2} + \Gamma_{\tau}^{2} } + \dfrac{\Gamma_{\tau}^{2} }{ \left(t - t_{0}^{(+)} \right)^{2} + \Gamma_{\tau}^{2} } \right\} ,
\label{eq39}
\end{equation}
\ \\
\noindent
peaks at $t = t_{0}^{(\mp)}$ when the particles are emitted.
Then from Eq.~(\ref{eq34_1}) we find to leading order in $\omega\Gamma_{\tau} \ll 1$,

\begin{equation}
\label{eq40}
I(t) \,=\, \frac{e}{\pi}\, \left\{ \dfrac{ \Gamma_{\tau} }{ \left( t - t_{0}^{(-)} \right)^2 + \Gamma_{\tau}^{2} } \,-\, \dfrac{\Gamma_{\tau} }{ \left( t - t_{0}^{(+)} \right)^2 + \Gamma_{\tau}^{2} } \right\} \,.
\end{equation}
\ \\
\noindent
This current consists of two pulses of Lorentzian shape with width $\Gamma_{\tau}$ corresponding to emission of an electron and a hole.
Integrating over time, it is easy to check that the first pulse carries a charge $e$ while the second pulse carries a charge $-e$.

The emitted particles carry energy from the dynamical localized state to the extended state and further to the reservoir which this extended state flows to.
The energy carried by the particles emitted during the period defines a heat generation rate $I_{E}$.
Substituting Eq.~(\ref{eq40}) into Eq.~(\ref{eq37}), we calculate

\begin{equation}
{\cal T} I_{E} \,=\, \dfrac{\hbar }{\Gamma_{\tau} }\,.
\label{eq41}
\end{equation}
\ \\
\noindent
Since there are two particles emitted during the period ${\cal T}$, the emitted particle (either an electron or a hole) carries an additional energy $\hbar/(2\Gamma_{\tau} )$ over the Fermi energy.
Since $I_{E} \ne 0$ the emitted particles are non-equilibrium particles.

\subsection{Multi-particle emission from
multiple localized states}\label{sec1.12}

\begin{figure}[t]
\centerline{\psfig{file=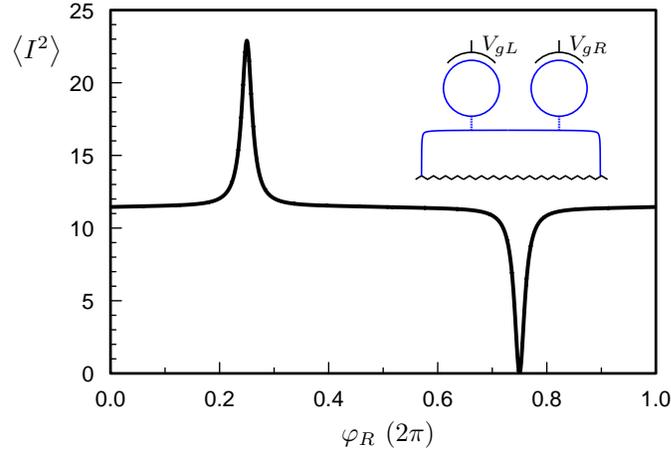,width=8.0cm}}
\begin{picture}(0,0)(0,0)  \put(170,0){$\varphi_{R}~(2\pi)$} \put(45,145){$\left \langle  I^{2} \right \rangle$} \put(223,148){\footnotesize$V_{gL}$} \put(256,148){\footnotesize$V_{gR}$ } \end{picture}
\caption{Inset: Two localized states (two edge states) coupled to the same extended state and reservoir. Main: The mean square current $\left \langle I^2  \right \rangle$ in units of $(e^2 \omega U_{j1} /\Delta )^2$ as a function of the phase $\varphi_{R}$. The parameters are: $\omega\Gamma_{j} = 1/(20\pi)$, $eU_{j1} = 0.5\Delta$, $\Delta_{j} = \Delta$ ($j = L\,,R$), $\varphi_{L} = \pi/2$.}
\label{fig1.3}
\end{figure}

If several localized states are placed in series along the same extended state (see inset to Fig.~\ref{fig1.3}), then such a combined structure can act as a multi-particle emitter.
Let the corresponding gates $V_{gL}$ and $V_{gR}$ induce the potentials $U_{j}(t) = U_{j0} + U_{j1}\, \cos\left( \omega t + \varphi_{j} \right)$, $j = L\,,R$, on the respective capacitors.
Then at times $t_{0j}^{(\zeta)}$ the cavity $j$ emits an electron ($\zeta =\, '-'$) and a hole ($\zeta =\, '+'$).
Since the emission times $t_{0j}^{(\zeta)}$ are defined by $\varphi_{j}$, then depending on the phase difference $\varDelta\varphi = \varphi_{L} - \varphi_{R}$ between the potentials $U_{L}(t)$ and $U_{R}(t)$ such a double-cavity capacitor can emit electron and hole pairs, or electron-hole pairs, or emit single particles, electrons and holes.

To recognize the emission regime it is convenient to analyze the mean square current generated by this structure,\cite{SOMB08}

\begin{eqnarray}
\langle I^{2} \rangle & = & \lim_{\varDelta t \to \infty}\,\frac{1 }{\varDelta t } \int\limits_{0}^{\varDelta t} dt\, I^{2}(t) \, \equiv \, \int\limits_{0}^{\cal T} \frac{dt}{\cal T}\, I^{2}(t)  \,.
\label{eq42}
\end{eqnarray}
\ \\
\noindent
In leading order in $\omega$ the current $I(t)$ produced by the coupled localized states is a sum of currents generated by each localized state separately, $I(t) = I_{L}(t) + I_{R}(t)$, where $I_{j}(t)$ is defined by Eq.~(\ref{eq40}) with $t_{0}^{(\mp)}$ and $\Gamma_{\tau}$ being replaced by $t_{0 j}^{(\mp)}$ and $\Gamma_{\tau j}$, correspondingly.
The advantage of considering $\langle I^{2} \rangle$ rather than $I(t)$ is the average over a long time instead of a more complicated time-resolved measurement.

Introducing the difference of times, $\varDelta t_{L,R}^{(\zeta, \zeta^\prime)} = t_{0L}^{(\zeta)} - t_{0R}^{(\zeta^\prime)}$, we find to leading order in $\omega\Gamma_{\tau j} \ll 1$,

\begin{eqnarray}
\langle I^{2}\rangle &=& \frac{e^2}{\pi {\cal T}  } \left( \frac{1}{\Gamma_{\tau L}} + \frac{1}{\Gamma_{\tau R}} \right)  \nonumber \\
\label{eq43} \\
&&\times  \bigg\{1 - L\left(\varDelta t_{L,R}^{(-,+)} \right) - L\left(\varDelta t_{L,R}^{(+,-)} \right) + L\left(\varDelta t_{L,R}^{(-,-)} \right) + L\left(\varDelta t_{L,R}^{(+,+)} \right) \bigg\} \,, \nonumber
\end{eqnarray}
\ \\
\noindent
where $L(\varDelta t) = 2\Gamma_{\tau L }\Gamma_{\tau R } \left\{ \left(\varDelta t\right)^{2} + \left( \Gamma_{\tau L} + \Gamma_{\tau R} \right)^{2} \right\}^{-1}$.
If both cavities emit particles at different times, $\left| \varDelta t_{L, R}^{(\zeta, \zeta^{\prime} )} \right| \gg \Gamma_{\tau j}$, then they contribute to the mean square current additively, $\langle I^{2}\rangle_{0} = e^2/(\pi {\cal T}) \left(\Gamma_{\tau L}^{-1} + \Gamma_{\tau R}^{-1} \right)$.
Below we use this quantity as a reference point.

Changing the phase lag $\varDelta\varphi$ one can enter the regime when one cavity emits an electron (a hole) at the time when the other cavity emits a hole (an electron), $\left| \varDelta t_{L, R}^{(-,+ )} \right| \lesssim \Gamma_{\tau j}$  $\left(\left| \varDelta t_{L, R}^{(+,- )} \right| \lesssim \Gamma_{\tau j}\right)$.
We expect that the source comprising both cavities does not generate a current, since the particle emitted by the first ($L$) cavity is absorbed by the second ($R$)  cavity.
Indeed, in this regime $\langle I^{2}\rangle = \langle I^{2}\rangle_{0} \left\{ 1 - L\left(\varDelta t_{L, R}^{(-,+ )} \right) - L\left(\varDelta t_{L, R}^{(+,- )} \right) \right\} $ is reduced indicating a reabsorption regime.\index{reabsorption regime}
For identical cavities, $\Gamma_{\tau L} = \Gamma_{\tau R}$, emitting in synchronism, $\varDelta t_{L,R}^{(-,+)} = \varDelta t_{L,R}^{(+,-)} = 0$, the mean square current vanishes, $\langle I^{2}\rangle = 0$. In Fig.~\ref{fig1.3} this synchronized regime shows up as a dip in the mean squared current.
Therefore, in this case the second cavity re-absorbs all the particles emitted by the first cavity.

An additional regime when the two particles of the same kind are emitted near simultaneously, $\left| \varDelta t_{L, R}^{(\zeta,\zeta )} \right| \lesssim \Gamma_{\tau j}$ is interesting.
In this case the mean square current is enhanced and for identical cavities emitting in synchronism it is, $\langle I^{2}\rangle = 2 \langle I^{2}\rangle_{0}$, and shows up as a peak in Fig.~\ref{fig1.3}.
The enhancement of the mean square current can be explained from the  energy perspective.
If two electrons (holes) are emitted simultaneously, then due to the Pauli exclusion principle they should be in different states, and that for spinless particles means different energies.
The second emitted particle should have an energy larger than the first one.
To be more precise, the electron (hole) pair has a larger energy then the sum of energies of two separately emitted electrons (holes).
Therefore, the heat flow $I_{E}$ generated by the double-cavity capacitor should be enhanced, that, by virtue of Eq.~(\ref{eq37}), implies an enhanced mean square current.

\section{Conclusion}\label{sec1.13}

The success of scattering theory of electrical transport for conductance and noise is well known.
We have shown that interesting conductance problems, of a qualitatively different nature, like the movement of charge in and out of a conductor, the ac-response to small amplitude voltage oscillations and even the large amplitude non-linear response find an elegant formulation in scattering theory. Novel
experiments in a 2D electron gas in the integer quantum Hall effect regime demonstrate\cite{gabelli06} quantization of the charge relaxation resistance of a mesoscopic cavity at half a von Klitzing resistance quantum. A successor experiment demonstrates\cite{feve07}
that a localized state can serve as a sub-nanosecond, single-electron source for coherent quantum electronics. Both of these phenomena seem to be well descried by scattering theory.
We only mention a few additional elements of a coherent quantum electronics using such sources in this paper.
The shot noise produced by the particles emitted by a dynamical localized state due to scattering at a quantum point contact is quantized, \index{quantized!shot noise} i.e., it is proportional to the number of emitted particles during an oscillation period ${\cal T}$.\cite{OSMB08,KSL08}
If two emitters are placed at different sides of a QPC then they contribute to shot noise additively, unless they are synchronized such that both cavities emit electrons (holes) at the same time.
In this case the shot noise vanishes.\cite{OSMB08}
This effect arises due to Fermi-correlations between electrons (holes) colliding at the QPC and propagating to different contacts.
This effect looks similar to the Hong, Ou, and Mandel\cite{HOM87} effect in optics.
However for fermions (electrons) the two-particle probability peaks while for bosons (photons) it shows a dip. \index{two-particle probability}
In collaboration with J.~Splettstoesser, we demonstrated that synchronized sources of uncorrelated particles can produce orbitally entangled pairs of electrons (holes).\cite{SMB09}
The amount of entanglement can be varied from zero to a maximum by a simple variation of the difference of phase $\varDelta\varphi$ of the potentials driving the localized states.

\section*{Acknowledgments}

We are grateful to Simon Nigg for discussion and technical assistance. We thank Janine Splettstoesser and Genevi{\`e}ve Fleury for a careful reading of the manuscript. This work is supported by the Swiss NSF, MaNEP, and by the European Marie Curie ITN NanoCTM.

\printindex[aindx]
\printindex

\end{document}